\documentclass[aps,prd,onecolumn,showpacs,nofootinbib,groupedaddress]{revtex4}  
\usepackage{graphicx}
\usepackage{epstopdf}
\usepackage{amsmath}
\usepackage{amsfonts}
\usepackage{amssymb}
\usepackage{appendix}
\usepackage{comment}
\usepackage{bbold}
\usepackage{color}
\usepackage{slashed}
\usepackage{subfigure}
\usepackage{setspace}
\usepackage{footnote}
\usepackage{multirow}
\usepackage{longtable}
\usepackage{braket}

\begin{document}

\singlespacing

{\hfill NUHEP-TH/17-06}

\title{Neutrino vs.~Antineutrino Oscillation Parameters at DUNE and Hyper-Kamiokande}

\author{Andr\'{e} de Gouv\^{e}a} 
\affiliation{Northwestern University, Department of Physics \& Astronomy, 2145 Sheridan Road, Evanston, IL 60208, USA}
\author{Kevin J. Kelly}
\affiliation{Northwestern University, Department of Physics \& Astronomy, 2145 Sheridan Road, Evanston, IL 60208, USA}

\begin{abstract}
Testing, in a non-trivial, model-independent way, the hypothesis that the three-massive-neutrinos paradigm properly describes nature is among the main goals of the current and the next generation of neutrino oscillation experiments. In the coming decade, the DUNE and Hyper-Kamiokande experiments will be able to study the oscillation of both neutrinos and antineutrinos with unprecedented precision. We explore the ability of these experiments, and combinations of them, to determine whether the parameters that govern these oscillations are the same for neutrinos and antineutrinos, as prescribed by the CPT-theorem. We find that both DUNE and Hyper-Kamiokande will be sensitive to unexplored levels of leptonic CPT-violation. Assuming the parameters for neutrino and antineutrino oscillations are unrelated, we discuss the ability of these experiments to determine the neutrino and antineutrino mass-hierarchies, atmospheric-mixing octants, and CP-odd phases, three key milestones of the experimental neutrino physics program. Additionally, if the CPT-theorem is violated in nature in a way that is consistent with all present neutrino and antineutrino oscillation data, we find that DUNE  and Hyper-Kamiokande have the potential to ultimately establish leptonic CPT-invariance violation.
\end{abstract}


\maketitle

\setcounter{equation}{0}
\section{Introduction}
\label{sec:introduction}

All neutrino oscillation data, with a few renowned exceptions, can be explained by hypothesizing  that neutrinos have nonzero, distinct masses and allowing for flavor-mixing in the charged-current weak interactions, as in the quark sector. It is remarkable that this three-massive-neutrinos paradigm, which adds to the Standard Model of Particle Physics half a dozen new parameters -- two independent neutrino mass-squared differences $\Delta m^2_{21}$ and $\Delta m^2_{31}$, three mixing angles $\theta_{12}, \theta_{13}, \theta_{23}$, and one CP-odd phase $\delta_{CP}$ \cite{Patrignani:2016xqp} -- provides an excellent fit to the data of more than a dozen very different neutrino experiments,  and that the values of many of these neutrino oscillation parameters are currently known at the several percent level \cite{Esteban:2016qun,deSalas:2017kay}. 

In spite of its extraordinary success, non-trivial ``over-constraining tests'' of the three-massive-neutrinos paradigm are still, for the most part, absent. Indeed, it is among the main goals of the current and the next generation of neutrino oscillation experiments to determine whether the three-massive-neutrinos paradigm properly describes nature or whether more leptonic new physics is required.

There are several candidates for the physics that might lie beyond the  three-massive-neutrinos paradigm. These incude the existence of new neutrino states or the existence of new weaker-than-weak neutrino--matter interactions. Here we investigate a more elementary question: to what extent do we know that neutrino and antineutrino oscillation parameters are the same, and how well will the next-generation long-baseline oscillation experiments -- Hyper-Kamiokande (Hyper-K)~\cite{Abe:2015zbg}  and the Deep Underground Neutrino Experiment (DUNE) ~\cite{Acciarri:2015uup} -- constrain this most reasonable hypothesis?

That particles and their antiparticles have equal masses and that their different couplings are intimately related is a consequence of the CPT-theorem. Comparing neutrino and antineutrino oscillation parameters amounts, therefore, to a particular test of CPT-conservation. Neutrino oscillation experiments can and have been used to perform different tests of the CPT-theorem \cite{Colladay:1996iz,deGouvea:2002xp,Bahcall:2002ia,Kostelecky:2003xn,Kostelecky:2003cr,Datta:2003dg,Minakata:2005jy,deGouvea:2006qd,Diaz:2009qk,Diaz:2011ia,Kostelecky:2011gq,Diaz:2013iba,Ohlsson:2014cha,Abe:2014wla,Arguelles:2015dca,Diaz:2016fqd} and we have nothing to add to these in this manuscript. It is challenging to construct simple models where particles and their antiparticles have different masses (for interesting attempts see, for example, \cite{Barenboim:2002tz,Chaichian:2012bk,Chaichian:2012ga,Fujikawa:2016her}) and it is fair to say that a detailed understanding of all the consequences of such scenarios is still the subject of active theoretical inquiry. Here, we will have nothing to add to these discussions and remain agnostic about the possibility that reasonable quantum-field-theory-like models that describe particles and antiparticles with different masses exist. 

Different oscillation parameters for neutrinos and antineutrinos have been postulated in the past in order to address discrepancies and disagreements in neutrino oscillation data \cite{Murayama:2000hm,Barenboim:2001ac}. While these proposals were quickly excluded by new experimental data or more detailed experimental data analyses \cite{GonzalezGarcia:2003jq}, it is well known that the current data allow for very large (order 100\%) effects \cite{deGouvea:2004va,Giunti:2010zs,Esteban:2016qun}. The situation is not expected to change qualitatively until the next generation of experiments starts taking data in the next decade. 
 
This manuscript is organized as follows. In Sec.~\ref{sec:Osc}, we properly define the hypothesis under investigation and summarize in a semi-quantitative way what existing data have to say about neutrino and antineutrino oscillation parameters. In Sec.~\ref{sec:Analysis}, we spell out the details of our simulations of the DUNE~\cite{Acciarri:2015uup} and Hyper-K~\cite{Abe:2015zbg} experiments, along with the accompanying beams from the Long-Baseline Neutrino Facility (LBNF) and the Japan Proton Accelerator Research Complex (J-PARC), respectively. We also discuss our treatment of atmospheric neutrino data in Hyper-K. In  Sec.~\ref{sec:Results}, we present and discuss our results. More concretely, we explore two types of scenarios. One assumes that nature is indeed CPT-conserving; there we address how well DUNE and Hyper-K can constrain CPT-violating oscillation parameters. The other assumes that nature violates CPT in the neutrino sector in a way that is consistent with current neutrino data; there we address whether DUNE and Hyper-K can establish that the CPT-theorem is indeed violated. In Sec.~\ref{sec:Conclusions}, we offer some closing remarks.

\setcounter{equation}{0}
\section{Neutrino and Antineutrino Oscillations}
\label{sec:Osc}

In this section, we will briefly introduce the notation used to describe neutrino and antineutrino oscillations and discuss the current measurements and constraints on the parameters of interest.  We make the assumption that only three neutrinos exist and that both neutrino and antineutrino interactions are as prescribed by the Standard Model. We allow, however, neutrino and antineutrino oscillation parameters -- mass-squared differences ($\Delta m_{ji}^2 \equiv m_j^2 - m_i^2$; $i,j=1,2,3$) and mixing matrices -- to be different. Throughout, we will refer to parameters $\vartheta$ associated to neutrino oscillations  using the standard notation ($\sin^2\theta_{13}$, $\Delta m^2_{21}$, etc) while antineutrino oscillation parameters $\overline{\vartheta}$ are identified with a bar ($\sin^2\overline{\theta}_{13}$, $\Delta \overline{m}^2_{21}$, etc). When discussing differences between neutrino and antineutrino oscillation parameters, we use $\Delta (\vartheta) \equiv \vartheta - \overline{\vartheta}$.

We make use of the Particle Data Group convention for the neutrino and antineutrino mixing matrixes $U$ and $\overline{U}$ \cite{Patrignani:2016xqp} and calculate oscillation probabilities using the standard formalism and all of the standard assumptions. Note that CPT-conservation translates into $\delta_{CP}=\overline{\delta}_{CP}$. We include the interactions between the neutrinos and electrons, protons, and neutrons in the path of propagation and assume that these matter effects are unaffected by whatever physics is responsible for violating the CPT-theorem. The same is true for the interactions associated with neutrino production and detection.

We will also consider the oscillation of neutrinos that are produced in the atmosphere and travel to the Hyper-K detector, potentially through long segments of the Earth. The density profile of the Earth is assumed to be highest at the core (density near 13 g/cm$^3$) and smallest near the crust (density near 3 g/cm$^3$). We assume the matter density profile of the Earth to be isotropic and piecewise constant, following the Preliminary Reference Earth Model \cite{Dziewonski:1981xy}. More of the details are presented in Ref.~\cite{Kelly:2017kch}.


\subsection{Existing Measurements of Neutrino and Antineutrino Oscillation Parameters}

Many studies exist in the literature highlighting and combining the most stringent measurements of and constraints on the neutrino oscillation parameters discussed above (the most recent ones are Refs.~\cite{Esteban:2016qun,deSalas:2017kay}). The vast majority of these analyses, of course, take advantage of the assumption that CPT is a good symmetry. Here, we review some of these measurements and estimate how well neutrino and antineutrino oscillation parameters are independently constrained. Our goal is not to perform a global fit to the world's neutrino oscillation data assuming neutrino and antineutrino oscillation parameters are independent. It turns out, however, that many parameters (or subsets of parameters) are predominantly constrained each by one experimental data set. In these cases, our estimates are a very good approximation for the result one would obtain when performing a global fit to the world's neutrino oscillation data. In later sections, when simulating the next-generation long-baseline neutrino oscillation experiments, we will use some of these estimates as prior information for the analyses that will follow.

\subsubsection{Antineutrino Oscillation Parameters}
\label{subsubsec:NuBarParams}

Antineutrino oscillation parameters are mostly constrained by reactor neutrino experiments \cite{Abe:2008aa,An:2016ses,RENO:2015ksa,Abe:2014bwa}, the Super-Kamiokande atmospheric neutrino data~\cite{Abe:2011ph}, and accelerator data from the Main Injector Neutrino Oscillation Search (MINOS) \cite{Adamson:2013whj} and the Tokai to Kamioka Experiment (T2K) \cite{Abe:2017bay}. These imply the existence of two hierarchical mass-squared differences $\Delta \overline{m}^2_{21}$ and $\Delta \overline{m}^2_{31}$ and three nonzero mixing angles $\overline{\theta}_{13}, \overline{\theta}_{12}, \overline{\theta}_{23}$. There is virtually no information regarding the CP-odd parameter $\overline{\delta}_{CP}$.

\textbf{KamLAND --} The reactor experiment KamLAND measured the survival probability of reactor electron-type antineutrinos after these have travelled distances around 100~km. Given independent information on $\overline{\theta}_{13}$ and $\Delta \overline{m}^2_{32}$, KamLAND data translates into measurements of $\Delta \overline{m}^2_{21} = 7.58^{+0.14}_{-0.13}$(stat)$^{+0.15}_{-0.15}$(syst)$\times 10^{-5}$ eV$^2$ and $\tan^2\overline{\theta}_{12} = 0.56^{+0.10}_{-0.07}$(stat)$^{+0.10}_{-0.06}$(syst) \cite{Abe:2008aa}. The collaboration states that this confidence interval for $\tan^2\overline{\theta}_{12}$ is obtained for $\overline{\theta}_{12}$ values in the first octant, $\overline{\theta}_{12}\in [0,\pi/4]$. In order to allow for values of $\overline{\theta}_{12}>\pi/4$ -- KamLAND cannot distinguish $\tan^2\overline{\theta}_{12}$ from $\cot^2\overline{\theta}_{12}$ -- we reinterpret this into a bound on $\sin^2{(2\overline{\theta}_{12})}$. We will make use of the following priors on these antineutrino oscillation parameters, whenever applicable:
\begin{align}
&\Delta \overline{m}_{21}^2 = \left(7.58 \pm 0.21\right) \times 10^{-5}\ \mathrm{eV}^2, \label{eq:antiDm12} \\
&\sin^2{(2\overline{\theta}_{12})} = 0.90 \pm 0.06,\label{eq:antiT12}
\end{align}
where we combined statistical and systematic errors in quadrature. 

\textbf{MINOS and T2K --} Accelerator experiments sensitive to $\bar{\nu}_{\mu}$-disappearance for $L/E$ values around $10^{3}$~eV$^{-2}$  provide the most stringent bounds on $\Delta \overline{m}^2_{32}$ and $\sin^2\overline{\theta}_{23}$ \cite{Adamson:2013whj,Abe:2017bay}. The results recently published by T2K are, for a normal mass-hierarchy, $\Delta \overline{m}^2_{32}=2.55^{+0.33}_{-0.27}\times 10^{-3}$~eV$^2$ and $\sin^2\overline{\theta}_{23}=0.42^{+0.25}_{-0.07}$ \cite{Abe:2017bay}. These are consistent with and roughly as precise as those from MINOS \cite{Adamson:2013whj} and atmospheric neutrino data~\cite{Abe:2011ph}. While analyzing simulated data from DUNE or Hyper-K, we will not include any prior information on $\Delta \overline{m}^2_{32}$ and $\sin^2\overline{\theta}_{23}$ since these are expected to be better constrained by DUNE and Hyper-K independently.

\textbf{Daya Bay, RENO, and Double Chooz --} Daya Bay \cite{An:2016ses}, RENO \cite{RENO:2015ksa}, and Double Chooz \cite{Abe:2014bwa} measure the survival probability of reactor electron-type antineutrinos $P_{\bar{e} \bar{e}}$ at baselines  around 1~km. Given independent information on $\overline{\theta}_{12}$ and $\Delta\overline{m}^2_{21}$,  they are sensitive to the mass-squared difference $\Delta \overline{m}_{32}^2$ and the mixing angle $\overline{\theta}_{13}$, and the most precise results were obtained with the Daya Bay experiment. Assuming a normal (inverted) mass-hierarchy, Daya Bay measures $\Delta \overline{m}_{32}^2  =\left(2.45 \pm 0.09\right) \times 10^{-3}$ eV$^2$ ($\Delta \overline{m}_{32}^2  =\left(-2.56 \pm 0.09\right) \times 10^{-3}$ eV$^2$). We expect, however, DUNE and Hyper-K to be more sensitive to this parameter, so we do not include this result as a prior in our analysis. We do include the most precise measurement of $\overline{\theta}_{13}$ \cite{An:2016ses},
\begin{align}
\sin^2{(2\overline{\theta}_{13})} = 0.0841 \pm 0.0033,
\label{eq:DB}
\end{align}
where we have combined statistical and systematic uncertainties in quadrature.

\subsubsection{Neutrino Parameter Measurements}
\label{subsubsec:NuParams}

Neutrino oscillation parameters are mostly constrained by solar neutrino experiments~\cite{Hampel:1997fc,Altmann:2005ix,Abdurashitov:2009tn,Aharmim:2011vm,Bellini:2013lnn,Abe:2016nxk}, the Super-Kamiokande atmospheric neutrino data~\cite{Abe:2011ph}, and accelerator data from MINOS \cite{Adamson:2014vgd}, T2K \cite{Abe:2017vif}, and the NuMI Off-Axis $\nu_e$  Appearance Experiment  (NO$\nu$A) \cite{Adamson:2017gxd}. These imply the existence of two hierarchical mass-squared differences $\Delta m^2_{21}$ and $\Delta m^2_{32}$ and three nonzero mixing angles $\theta_{13}, \theta_{12}, \theta_{23}$. There is very limited information regarding the CP-odd parameter $\delta_{CP}$.

\textbf{Solar Neutrinos --} Given everything else that is known about neutrino oscillations, solar neutrino data are most sensitive to the neutrino parameters $\Delta m^2_{21}$ and $\sin^2\theta_{12}$. They also provide limited information on $\sin^2\theta_{13}$, which we will not discuss here. The NuFIT collaboration \cite{Esteban:2016qun} combines results from the Sudbury Neutrino Observatory (SNO), Super-Kamiokande, and Borexino experiments in a global analysis, obtaining $\Delta m_{21}^2 = (5.4 \pm 1.0) \times 10^{-5}$~eV$^2$. This preferred value of $\Delta m_{21}^2$ is smaller than the preferred value of $\Delta \overline{m}^2_{21}$ obtained by the KamLAND experiment (see above), with a roughly $2\sigma$ disagreement between the two measurements. 

On the other hand, in combined analyses of solar neutrino data and the data from the KamLAND experiment, constraints on $\tan^2{\theta_{12}}$ are dominated by solar data, being mostly insensitive to KamLAND data. For this reason, as far as $\theta_{12}$ is concerned, we simply take the results from NuFIT \cite{Esteban:2016qun}. We highlight that the solar neutrino measurements, in the absence of new interactions, constrain $\theta_{12}$ to be in the first octant, so expressing results in terms of $\tan^2\theta_{12}$ or $\sin^2\theta_{12}$ is appropriate here. In summary, we will make use of the following priors on these neutrino oscillation parameters, whenever applicable:
\begin{align}
&\Delta m_{21}^2 = \left( 5.4 \pm 1.0 \right) \times 10^{-5}\ \mathrm{eV}^2,\label{eq:Dm12} \\
&\tan^2\theta_{12} = 0.452 \pm 0.035.\label{eq:T12}
\end{align}

\textbf{MINOS, T2K and NO$\nu$A--} Accelerator experiments sensitive to $\nu_{\mu}$-disappearance for $L/E$ values around $10^{3}$~eV$^{-2}$  provide the most stringent bounds on $\Delta m^2_{32}$ and $\sin^2\theta_{23}$ \cite{Adamson:2014vgd,Abe:2017bay,Abe:2017vif,Adamson:2017gxd}. The results recently published by T2K are $\Delta m_{32}^2 = \left(2.54 \pm 0.08\right) \times 10^{-3}$ eV$^2$ and $\sin^2\theta_{23}=0.55^{+0.05}_{-0.09}$ \cite{Abe:2017vif}, both for a normal mass-hierarchy. Results for the inverted hierarchy are similar (up to $\Delta m^2_{32}$ approximately changing sign). These are consistent with results from MINOS \cite{Adamson:2014vgd}, NO$\nu$A \cite{Adamson:2017gxd}, and atmospheric neutrino data~\cite{Abe:2011ph}. While analyzing simulated data from DUNE or Hyper-K, we will not include any prior information on $\Delta m^2_{32}$ and $\sin^2\theta_{23}$ since these are expected to be better constrained by DUNE and Hyper-K independently. 

T2K and NO$\nu$A are also sensitive to the value of $\sin^2\theta_{13}$ \cite{Abe:2017vif,Adamson:2017gxd}. The T2K collaboration presented results on neutrino and antineutrino oscillation parameters separately, resulting in a bound on $\sin^2\theta_{13}$ between $0.02$ and $0.065$ at $68.3\%$ confidence level \cite{Abe:2017vif}. While analyzing simulated data from DUNE or Hyper-K, we will not include any prior information on  $\sin^2\theta_{13}$ since it is expected to be better constrained by DUNE and Hyper-K independently. Finally, it is well known that accelerator experiments, combined with information from the reactor experiments, already provide hints of CP violation, or $\delta_{CP} \neq 0$, $\pi$. However, these hints rely heavily on information from reactor antineutrino experiments and therefore assume CPT is a good symmetry. Here, we assume $\delta_{CP}$ is presently unconstrained.

\subsubsection{Summary of Current Constraints on CPT-Violating Oscillation Parameters}

It is interesting to appreciate that, given current neutrino data, we can affirm that both neutrinos and antineutrinos oscillate. Furthermore, there is unequivocal evidence that the description of both neutrino and antineutrino oscillations requires two different oscillation frequencies and that the neutrino and antineutrino oscillation frequencies are at least qualitatively the same. 

More quantitatively, $\Delta(\Delta m^2_{32})$ is the best constrained (relative to the average value) CPT-violating observable if one assumes the same mass-hierarchy for neutrinos and antineutrinos. Using T2K data from Ref.~\cite{Abe:2017bay}, we estimate, assuming a normal mass-hierarchy in both sectors, 
\begin{equation}
\Delta(\Delta m^2_{32}) = -(0.02\pm0.32) \times 10^{-3}~{\rm eV}^2,
\label{eq:DDm13}
\end{equation} 
constrained to be at most fifteen percent of the average value. Order 100\% CPT-violating effects are not excluded, however, since the neutrino and antineutrino mass orderings could be distinct (e.g., $\Delta m^2_{32}>0$, $\Delta \overline{m}^2_{32}<0$). The difference between the longer wavelength oscillation mass-squared-differences, however, is only poorly constrained (relative to the average value), 
\begin{equation}
\Delta(\Delta m^2_{21}) = -(2.18\pm1.02) \times 10^{-5}~{\rm eV}^2,
\label{eq:DDm12}
\end{equation}
and, as already mentioned, $\Delta(\Delta m^2_{21})$ is nonzero at around the 2$\sigma$ level. 

For the mixing matrices, the differences between neutrinos and antineutrinos are all order 100\%, albeit for different reasons. Both $\Delta(\theta_{12})$ and $\Delta(\theta_{23})$ are poorly constrained partially because of ``dark side'' possibilities \cite{deGouvea:2004va}. In the case of $\Delta(\theta_{12})$, solar data guarantee $\theta_{12}<\pi/4$, while KamLAND data, even if KamLAND were to run for a very long time and systematic effects were very small, cannot distinguish $\overline{\theta}_{12}$ from $\pi/2-\overline{\theta}_{12}$. $\Delta(\theta_{23})$ is poorly constrained because we could have $\theta_{23}<\pi/4$ and $\overline{\theta}_{23}>\pi/4$, or vice-versa, and because the current error bars on the individual parameters are larger than ten percent. On the other hand, $\Delta(\theta_{13})$ is poorly constrained because the measurements of $\theta_{13}$, as opposed to $\overline{\theta}_{13}$, are still very poor (order 100\% uncertainty). Needless to say, given how poorly the CP-violating parameters are individually constrained, any $-2\pi<\Delta({\delta}_{CP})<2\pi$ is allowed.\footnote{$\Delta(\delta_{CP})$ can be defined in a variety of ways. Here, we assume both $\delta_{CP},\overline{\delta}_{CP}\in[-\pi,\pi)$ so $\Delta(\delta_{CP})\in (-2\pi,2\pi)$.}

\setcounter{equation}{0}
\section{Analysis Method}
\label{sec:Analysis}
In order to explore the potential of next-generation experiments to test the CPT-theorem, we proceed as follows. For a given set of neutrino and antineutrino oscillation parameters, we compute the expected number of signal and background events as a function of energy for the experiment of interest, as outlined in more detail in \cite{Berryman:2015nua,deGouvea:2015ndi} for DUNE\footnote{Here, the expected signal and background events are updated relative to the analyses performed in Refs.~\cite{Berryman:2015nua,deGouvea:2015ndi} and are consistent with those from the most recent simulations published by the DUNE Collaboration \cite{Acciarri:2015uup}.} and Ref.~\cite{Kelly:2017kch} for Hyper-K. More concretely, we generate expected signal and background yields for the following experimental oscillation channels: 
\begin{itemize}
\item DUNE:
\begin{itemize}
\item electron-type neutrino appearance,
\item electron-type antineutrino appearance,
\item muon-type neutrino disappearance,
\item muon-type antineutrino disappearance.
\end{itemize}
\item Hyper-Kamiokande:
\begin{itemize}
\item electron-type neutrino appearance (beam-based),
\item electron-type antineutrino appearance (beam-based),
\item muon-type neutrino disappearance (beam-based),
\item muon-type antineutrino disappearance (beam-based),
\item atmospheric muon-type neutrino disappearance (low-energy),
\item atmospheric muon-type neutrino disappearance (high-energy).
\end{itemize}
\end{itemize}

Our simulation of the DUNE signal and background yields, in agreement with Ref.~\cite{Acciarri:2015uup}, corresponds to a total exposure of 300 kton-MW-years, which the collaboration claims corresponds to seven years of data collection with a $1.07$~MW, $80$ GeV proton beam. This includes equal time in neutrino and antineutrino modes.\footnote{Both the J-PARC and LBNF beams contain mixtures of muon-type neutrinos and antineutrinos and electron-type neutrinos and antineutrinos regardless of whether they are running in ``neutrino mode'' or ``antineutrino mode.'' In our analyses, of course, neutrinos oscillate according to the neutrino oscillation parameters and antineutrinos oscillate according to the antineutrino oscillation parameters, regardless of which beam mode the experiments are in.} Our simulation of Hyper-K, in agreement with Ref.~\cite{Abe:2015zbg}, assumes ten years of data collection with a $30$ GeV proton beam, amounting to a total of $1.56 \times 10^{22}$ protons on target. As opposed to DUNE, Hyper-K will spend more time collecting data in the antineutrino mode, with a run-time-ratio of $1:3$ neutrino to antineutrino mode. We simulate the atmospheric muon neutrino samples by inflating the atmospheric neutrino sample of Super-Kamiokande to correspond to ten years of data collection with a detector twenty times larger \cite{Kelly:2017kch}.

For parameter-estimation we make use of a chi-squared test statistic that is a function of twelve independent physics parameters -- six for neutrinos, six for antineutrinos. In order to include information from existing experiments, we add to the chi-squared (some of) the Gaussian priors -- Eqs.~(\ref{eq:antiDm12}) - (\ref{eq:T12}) -- discussed in Sections~\ref{subsubsec:NuBarParams} and \ref{subsubsec:NuParams}, as well as normalization uncertainties on the signal and background yields. We assume signal and background normalization uncertainties of 5\% each for both the DUNE and Hyper-K beam-based yields, and a 10\% normalization uncertainty for the Hyper-K atmospheric-based yields.

Using the Markov Chain Monte Carlo package {\sc emcee} \cite{ForemanMackey:2012ig}, we are able to estimate the posterior likelihood distributions of each parameter or pair of parameters. We consider various combinations of the oscillation channels listed above in order to probe CPT violation. These combinations include 
\begin{itemize}
\item the DUNE data set alone (DUNE), 
\item the Hyper-Kamiokande beam-based data set alone (HK B), 
\item the Hyper-Kamiokande atomspheric- and beam-based data sets (HK AB),  
\item the DUNE data set and Hyper-Kamiokande beam-based data set (DUNE + HK B), and 
\item the DUNE data set and Hyper-Kamiokande atmospheric- and beam-based data sets (DUNE + HK AB). 
\end{itemize}
We also perform some analyses  without including the Daya Bay prior on $\sin^2{\left(2\overline{\theta}_{13}\right)}$, Eq.~(\ref{eq:DB}), in order to see how it impacts our results. In these cases, all oscillation  parameters except for the solar/KamLAND-ones -- $\Delta m^2_{21}, \theta_{12}, \Delta \overline{m}^2_{21}, \overline{\theta}_{12}$ -- are being measured exclusively by the same combination of DUNE and Hyper-K simulated data.


\setcounter{equation}{0}
\section{Results}
\label{sec:Results}

In this section, we discuss the capabilities of DUNE and Hyper-K to test the CPT-theorem. In Section~\ref{subsec:Simultaneous}, we address how well the neutrino and antineutrino oscillation parameters can be measured independently from one another assuming CPT is a good symmetry. We highlight the $\theta_{23},\overline{\theta}_{23}$ octant degeneracy and whether it can be resolved if one chose to treat neutrino and antineutrino oscillation parameters as independent. In Section~\ref{subsec:Exclusion}, we estimate the ability of next-generation experiments to constrain CPT-violation, assuming CPT is a good symmetry. In Section~\ref{subsec:MeasViolation}, we simulate data under the assumption that CPT is violated (in two different ways), and analyze how well this violation would be measured by various combinations of DUNE and Hyper-Kamiokande data.

\subsection{Independent Measurement of Neutrino and Antineutrino Parameters}
\label{subsec:Simultaneous}

Here we simulate data consistent with the CPT-theorem. We use as input for the neutrino and antineutrino oscillation parameters the global-fit results presented in Ref.~\cite{Esteban:2016qun}, except for $\sin^2\theta_{23}$. These inputs are tabulated in Table~\ref{tab:InputParamsExclusion}. As far as $\sin^2\theta_{23}$ is concerned, we consider two different hypotheses: $\sin^2\theta_{23}=0.441$ (from the global-fit results presented in Ref.~\cite{Esteban:2016qun})  and $\sin^2\theta_{23}=0.500$ (``maximal mixing''). 
\begin{table}[!htbp]
\begin{center}
\begin{tabular}{|r||c|c|c|c|c|c|}
\hline
Parameter & $\sin^2\theta_{12}$ & $\sin^2\theta_{13}$ & $\sin^2\theta_{23}$ & $\delta_{CP}$ & $\Delta m_{21}^2$ & $\Delta m_{31}^2$ \\ \hline
Value & $0.306$ & $0.02166$ & $0.441$ or $0.500$ & $-1.728$ & $7.50 \times 10^{-5}$ eV$^2$ & $+2.524\times 10^{-3}$ eV$^2$ \\ \hline 
\end{tabular}
\end{center}
\caption{Input parameter values for the analyses performed in Sections~\ref{subsec:Simultaneous} and \ref{subsec:Exclusion}. All inputs -- which are the same for neutrinos and antineutrinos -- are taken from the NuFIT collaboration \cite{Esteban:2016qun}, except for $\sin^2\theta_{23}$, which is assumed to be maximal for portions of Section~\ref{subsec:Simultaneous} and all of Section~\ref{subsec:Exclusion}.}
\label{tab:InputParamsExclusion}
\end{table}

\begin{figure}[!htbp]
\begin{center}
\includegraphics[width=0.8\linewidth]{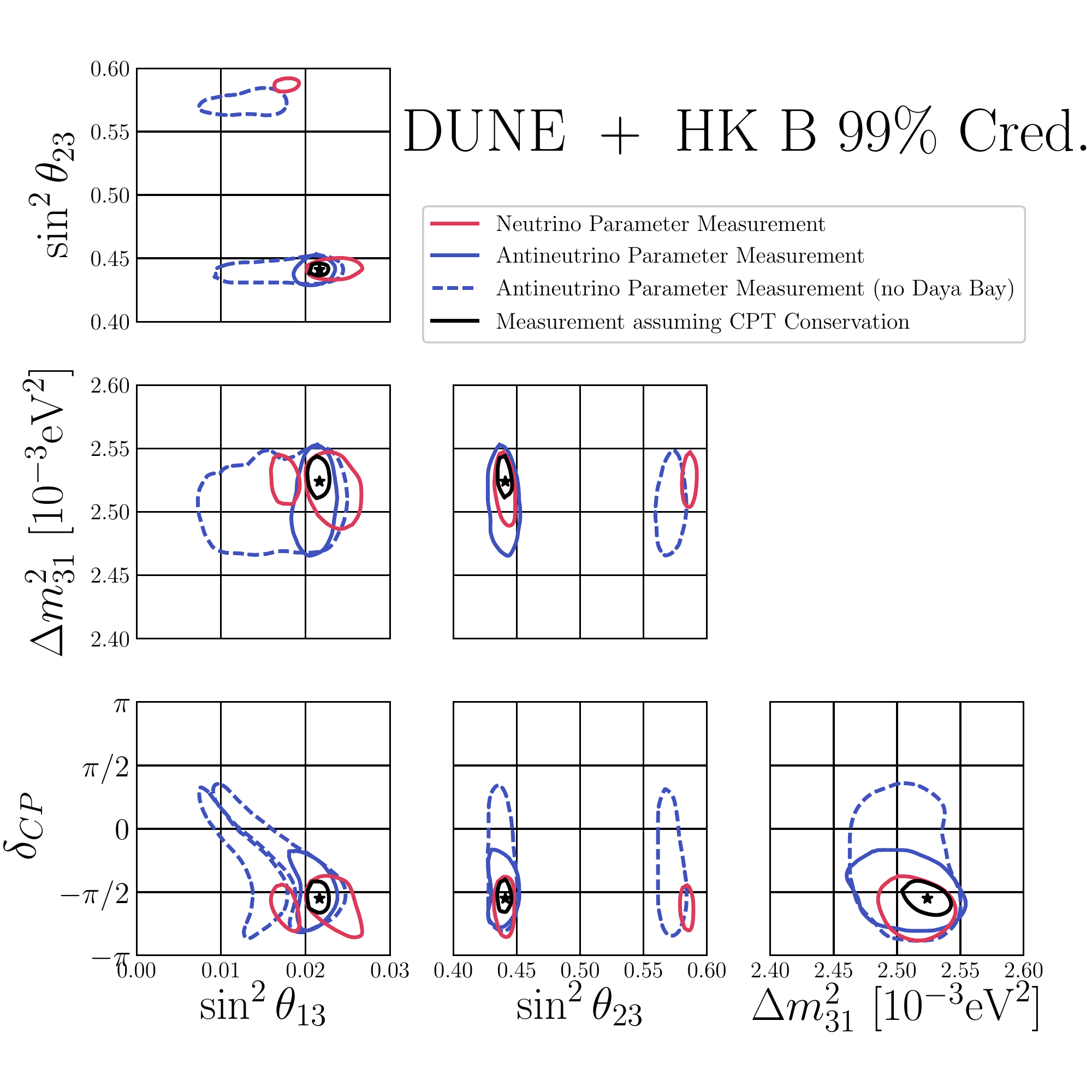}
\end{center}
\caption{Allowed regions of parameter space at 99\% credibility assuming the true values of parameters in Table~\ref{tab:InputParamsExclusion} with $\sin^2\theta_{23} = 0.441$. Parameters are projected down to common axes for comparison. The red contours display the regions allowed for the neutrino parameters, and blue contours display the regions allowed for antineutrino parameters (bars on parameters omitted). We also include the results of a CPT-conserving analysis in black. The dashed regions display the additional parameter space allowed at 99\% credibility when the analysis is repeated without the prior on $\sin^2(2\overline{\theta}_{13})$ from the Daya Bay experiment, Eq.~(\ref{eq:DB}).}
\label{fig:OverlayLow}
\end{figure}

\begin{figure}[!htbp]
\begin{center}
\includegraphics[width=0.8\linewidth]{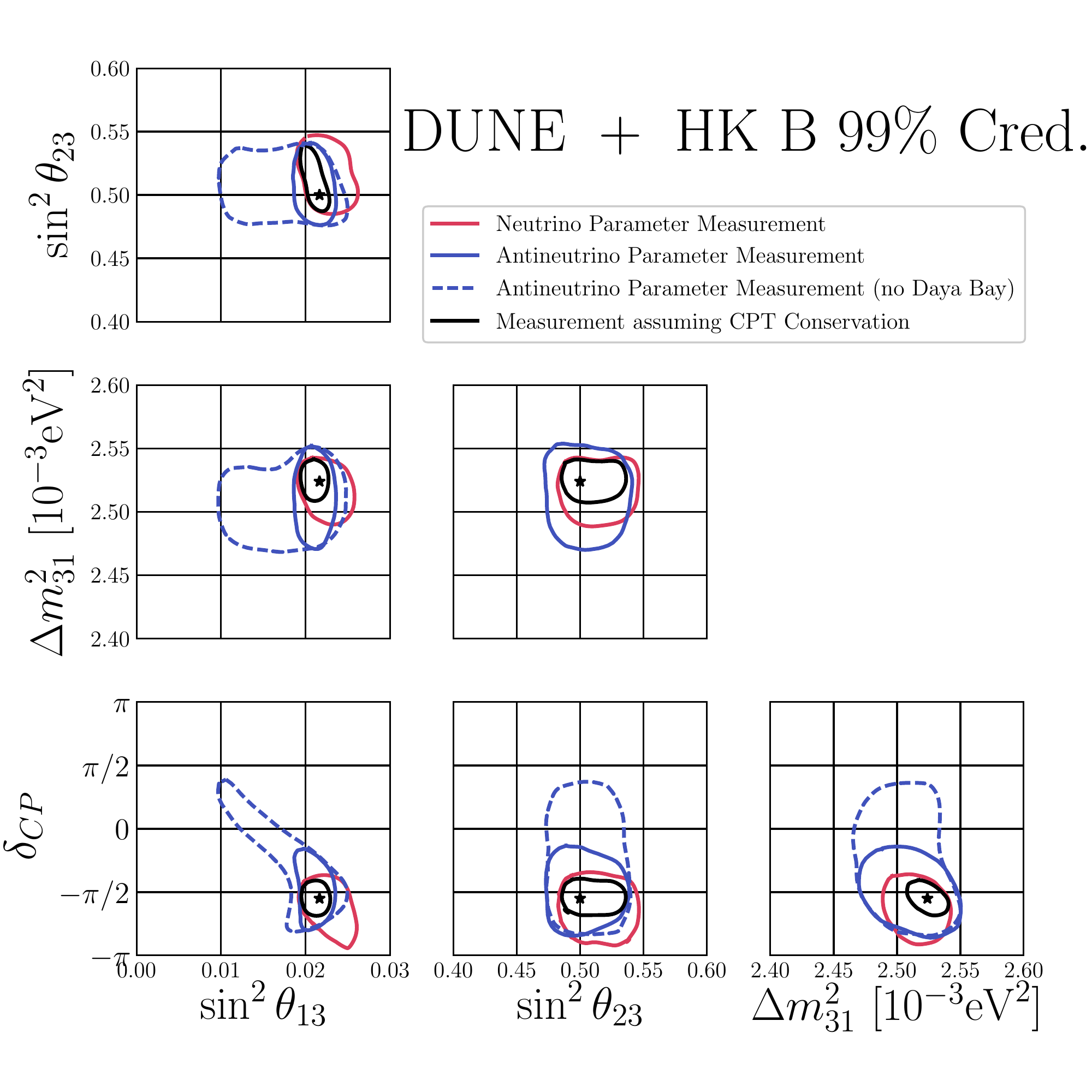}
\end{center}
\caption{Allowed regions of parameter space at 99\% credibility assuming the true values of parameters in Table~\ref{tab:InputParamsExclusion} with $\sin^2\theta_{23} = 0.500$. Parameters are projected down to common axes for comparison. The red contours display the regions allowed for the neutrino parameters, and blue contours display the regions allowed for antineutrino parameters (bars on parameters omitted). We also include the results of a CPT-conserving analysis in black. The dashed regions display the additional parameter space allowed at 99\% credibility when the analysis is repeated without the prior on $\sin^2(2\overline{\theta}_{13})$ from the Daya Bay experiment, Eq.~(\ref{eq:DB}).}

\label{fig:OverlayMax}
\end{figure}

Fig.~\ref{fig:OverlayLow} depicts the 99\% credible region for neutrino and antineutrino oscillation parameters, assuming $\sin^2\theta_{23} = 0.441$, obtained upon analyzing DUNE + HK B simulated data, defined in Sec.~\ref{sec:Analysis}. Fig.~\ref{fig:OverlayMax} depicts the same information, assuming $\sin^2\theta_{23} = 0.500$. While the axes are labelled using neutrino oscillation parameters, it is understood that these are to be read as antineutrino oscillation parameters when applicable. Solid red contours refer to neutrino oscillation parameters, while solid blue contours refer to antineutrino oscillation parameters. In the case of antineutrino oscillation parameters, we also include -- dashed blue contours -- the allowed regions one extracts without including prior information from the Daya Bay reactor experiment, Eq.~(\ref{eq:DB}). Solid black contours refer to the results we obtain by analyzing the data assuming the neutrino and antineutrino parameters are the same.\footnote{In this case, the number of physics parameters in the fit is six.} Note that these are joint analyses of DUNE + HK B and each panel in Figs.~\ref{fig:OverlayLow} and \ref{fig:OverlayMax} depicts results for two of the twelve parameters obtained upon marginalizing over the remaining ten parameters. The marginalization procedure includes the mass-hierarchy, both in the neutrino and antineutrino sectors. The figures do not depict the extracted values of $\Delta m^2_{21}, \sin^2\theta_{12}$ and the corresponding antineutrino ``KamLAND'' parameters. The reason is that DUNE and Hyper-K are only able to constrain these long-wavelength oscillation parameters marginally and virtually all information is provided by the existing data, captured by the priors discussed in Sec.~\ref{sec:Osc}.


Fig.~\ref{fig:OverlayLow} reveals that DUNE + HK B cannot resolve the octant degeneracy as far as the neutrino oscillation parameters are concerned. This unresolved degeneracy also leads to two distinct (at the 99\%~level) extracted values of $\sin^2\theta_{13}$. Without resorting to the CPT-theorem, this degeneracy probably cannot be resolved without the existence of a qualitatively more powerful long-baseline neutrino oscillation experiment. For antineutrino parameters, the situation is similar before existing reactor-antineutrino information on $\sin^2\overline{\theta}_{13}$ is included in the analysis. It is useful to remember that DUNE or Hyper-K, by themselves, can resolve the octant degeneracy if CPT-conservation is assumed \cite{Abe:2015zbg,Acciarri:2015uup}. 

Overall, as expected, neutrino parameters are constrained more stringently than antineutrino parameters. This is driven by the fact that there is more statistical power in the ``neutrino beams'' than in the ``antineutrino beams.'' Furthermore, ``neutrino beams'' are cleaner than ``antineutrino beams.'' If both $\delta_{CP}$ and $\overline{\delta}_{CP}$ are as large as what is listed in Table~\ref{tab:InputParamsExclusion}, both neutrino and antineutrino CP-violation can be independently established at 99\% credibility. In the case of antineutrino CP-violation, existing reactor-antineutrino information on $\sin^2\overline{\theta}_{13}$ plays a fundamental role.

Here we concentrated on the DUNE + HK B data sample and will not present results for other simulated data samples to avoid an overabundance of figures qualitatively similar to Figs.~\ref{fig:OverlayLow} and \ref{fig:OverlayMax}. Our main goal was to highlight the power and limitations of making oscillation measurements without assuming the neutrino and antineutrino oscillation parameters are the same. We explore the statistical power of the different data sets in the next subsections.

\subsection{Constraining CPT-Violation}
\label{subsec:Exclusion}

If the data are consistent with CPT-conservation, i.e., if the extracted values of the neutrino and antineutrino oscillation parameters comfortably agree, we can place bounds on CPT-violation. As in the previous subsection,  we simulate data assuming that CPT is conserved in nature, and analyze those data under the assumption that the parameters for neutrino and antineutrino oscillations are independent. The input values for these parameters are listed in Table~\ref{tab:InputParamsExclusion} but we restrict all analyses here to $\sin^2\theta_{23}=0.5$. We assume that the mass-hierarchy is normal, i.e. $\Delta m_{31}^2 > 0$, for both neutrinos and antineutrinos, but marginalize over all possible combinations of the mass-hierarchies when presenting results. 

While we assume that CPT is conserved in nature in this subsection, the solar/KamLAND priors discussed in Sections~\ref{subsubsec:NuBarParams} and ~\ref{subsubsec:NuParams} slightly disagree with the inputs. The contribution to the value of chi-squared from these priors given the input parameters in Table~\ref{tab:InputParamsExclusion} is $5.39$ -- a small shift that we take into account when estimating parameters. We further checked that the impact on the extraction of the bounds on CPT-violation described below is minimal.   

\begin{figure}[!htbp]
\begin{center}
\includegraphics[width=\linewidth]{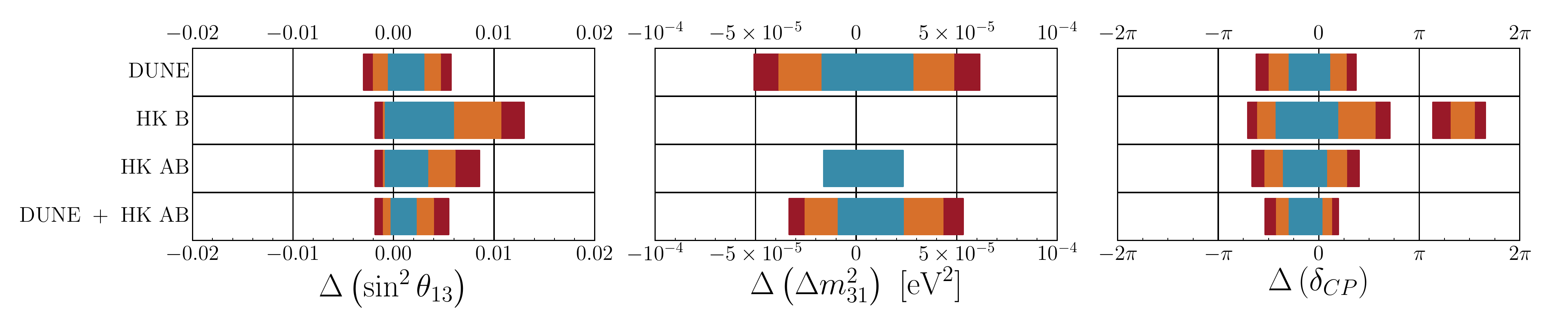}
\caption{Sensitivity to CPT violation at various combinations of experiments. Each row displays sensitivity to $\Delta (\sin^2\theta_{13})$ (left), $\Delta (\Delta m_{31}^2)$ (center), and $\Delta (\delta_{CP})$ (right) at 68.3\% (blue), 95\% (orange), and 99\% (red) credibility. Here, ``HK B'' represents the Hyper-Kamiokande beam-based events, and ``HK AB'' represents the combination of beam- and atmospheric-based events. In this figure, the prior information on $\sin^2(2\overline{\theta}_{13})$ from the Daya Bay experiment is included.}
\label{fig:ExclusionResultsDB}
\end{center}
\end{figure}

\begin{figure}[!htbp]
\begin{center}
\includegraphics[width=\linewidth]{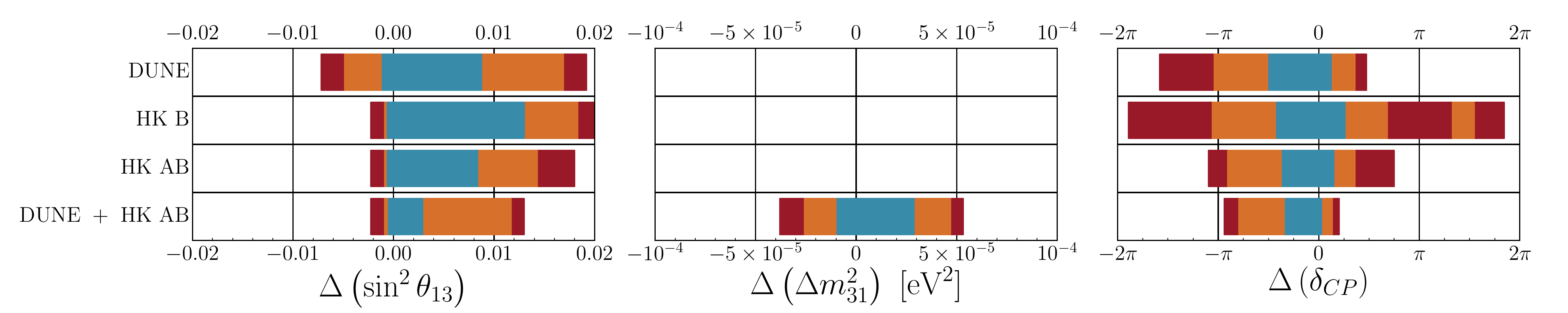}
\caption{Sensitivity to CPT violation at various combinations of experiments. Each row displays sensitivity to $\Delta (\sin^2\theta_{13})$ (left), $\Delta (\Delta m_{31}^2)$ (center), and $\Delta (\delta_{CP})$ (right) at 68.3\% (blue), 95\% (orange), and 99\% (red) credibility. Here, ``HK B'' represents the Hyper-Kamiokande beam-based events, and ``HK AB'' represents the combination of beam- and atmospheric-based events. In this figure, the prior information on $\sin^2(2\overline{\theta}_{13})$ from the Daya Bay experiment is not included.}
\label{fig:ExclusionResultsNoDB}
\end{center}
\end{figure}

We marginalize over all-but-one of the twelve different oscillation parameters and extract the probability distribution for $\Delta(\vartheta)$ for $\vartheta = \sin^2\theta_{13}$, $\Delta m_{31}^2$, and $\delta_{CP}$, and calculate the 68.3\%, 95\%, and 99\% credible ranges for each of these.  Figs.~\ref{fig:ExclusionResultsDB} and \ref{fig:ExclusionResultsNoDB} depict the results of this procedure for the different combinations of simulated data introduced in Section~\ref{sec:Analysis}. The data analysis that leads to Fig.~\ref{fig:ExclusionResultsDB} includes the prior on $\sin^2\overline{\theta}_{13}$ from Daya Bay, Eq.~(\ref{eq:DB}), not included in the analysis that leads to Fig.~\ref{fig:ExclusionResultsNoDB}.

DUNE by itself can constrain $|\Delta(\sin^2\theta_{13})|/\sin^2\theta_{13}$ at around the 30\% level, and the situation is slightly improved if one includes atmospheric and beam data from Hyper-K. This is only true with the help of existing Daya Bay data which, as illustrated in the previous subsection, is necessary in order to ``pin-point'' the value of $\sin^2\overline{\theta}_{13}$. Without existing reactor data, the best one can do with next-generation experiments is to constrain $|\Delta(\sin^2\theta_{13})|/\sin^2\theta_{13}$ to be smaller than around 50\%. Here there is a preference for $\Delta (\sin^2\theta_{13}) > 0$. This is due to the input-choice for $\delta_{CP}$, as well as the assumption that the mass-hierarchy is normal. The impact of $\delta_{CP}$ can be seen in Figs.~\ref{fig:OverlayLow} and~\ref{fig:OverlayMax} in the $\sin^2\theta_{13} \times \delta_{CP}$ plane, as the contours for neutrino and antineutrino parameters prefer to ``spread'' in opposite directions.

As for the CP-violating phase, DUNE can, by itself, constrain $\Delta(\delta_{CP})$ to be smaller than, roughly, $\pm \pi/2$, and the situation is improved if one includes atmospheric and beam data from Hyper-K. This means that, for the parameters of choice, both neutrino and antineutrino CP-violation would be independently established. Also here the prior on $\sin^2\overline{\theta}_{13}$ is crucial. Fig.~\ref{fig:ExclusionResultsNoDB} reveals that in the absence of precision measurements of $\sin^2\overline{\theta}_{13}$ from reactor experiments it is not possible to establish that CP is violated in the antineutrino sector, even if $\overline{\delta}_{CP}$ deviates maximally from 0 and $\pi$.

The bound on $\Delta(\Delta m^2_{31})$ is expected to ultimately improve by almost an order of magnitude relative to the existing bound, Eq.~(\ref{eq:DDm13}). As is the case today, a stringent bound is only possible if the data can independently establish that the neutrino and antineutrino mass-hierarchies are the same. Figs.~\ref{fig:ExclusionResultsDB} and \ref{fig:ExclusionResultsNoDB} illustrate this fact as follows. The DUNE experiment can determine the mass-hierarchy for both neutrinos and antineutrinos independently, but only if the Daya Bay prior is included. Hence a non-trivial bound is included in Fig.~\ref{fig:ExclusionResultsDB} but not in Fig.~\ref{fig:ExclusionResultsNoDB}. On the other hand, DUNE + HK AB can determine the mass-hierarchy for both neutrinos and antineutrinos independently, even if information from Daya Bay is ignored. We have verified that both these statements are independent of the input value of $\delta_{CP}$. HK B cannot determine the neutrino mass-hierarchy even if information from Daya Bay -- which helps resolving the antineutrino mass-hierarchy -- is included. For HK AB, the two mass-hierarchies can be determined independently only at the one-sigma level and only if the Daya Bay prior is included, and just for the value of $\delta_{CP}$ we've chosen. This is indicated by the solitary blue bar in Fig.~\ref{fig:ExclusionResultsDB}. In this case, the main source of ambiguity is the combination $\Delta m_{31}^2 < 0$, $\Delta \overline{m}_{31}^2 > 0$, which ``works'' for large values of $\sin^2\theta_{13}$. Additional constraints on $\sin^2\theta_{13}$ from, e.g., NO$\nu$A and T2K, could help HK AB determine the mass-hierarchies independently. We also draw attention to the fact we are only including the muon-type neutrinos and antineutrinos seen in the atmospheric data. Atmospheric data on electron-type neutrinos and antineutrinos should add invaluable information when it comes to determining the neutrino and antineutrino mass-hierarchies, thanks to large matter effects. 

Ultimately, magnitude-wise, we expect $\Delta(\Delta m^2_{31})$ to be as constrained as $\Delta(\Delta m^2_{21})$.\footnote{In the absence of new solar neutrino experiments, better CPT-violating bounds on the long-wavelength parameters are expected from the JUNO experiment \cite{An:2015jdp} (and also RENO-50 \cite{Kim:2014rfa}). The error bar in Eq.~(\ref{eq:DDm12}), however, is dominated by the solar neutrino data and is not expected to improve significantly. The central value, of course, could change, but only if future JUNO/RENO-50 data disagree with existing KamLAND data.} Given that $\Delta(\Delta m^2_{21})$ is currently $2\sigma$ away from zero, this may prove to be an important sensitivity level to reach.

Our analyses allow us to also display results for $\Delta(\sin^2\theta_{12}$), $\Delta(\sin^2\theta_{23})$, and $\Delta(\Delta m_{21}^2)$, similar to Figs.~\ref{fig:ExclusionResultsDB} and \ref{fig:ExclusionResultsNoDB}. As already mentioned, the results for $\Delta(\sin^2\theta_{12})$ and $\Delta(\Delta m_{21}^2)$ are driven completely by the priors on $\tan^2\theta_{12}$, $\sin^2{\left(2\overline{\theta}_{12}\right)}$, $\Delta m_{21}^2$, and $\Delta \overline{m}_{12}^2$, and neither DUNE nor Hyper-K are not expected to add very much to the discussion. We refrain from depicting $\Delta(\sin^2\theta_{23})$ for a different reason. The octant degeneracy, as we demonstrated in the previous subsection, cannot be lifted for the neutrino oscillation parameters. This means that, for non-maximal $\sin^2\theta_{23}$, $\Delta(\sin^2\theta_{23})\sim {\cal O}(1)$. On the other hand, for strictly maximal $\sin^2\theta_{23}$ the allowed region for $\Delta(\sin^2\theta_{23})$ is relatively small. Hence, in order to avoid giving the wrong impression about the ability of next-generation experiments to constrain $\Delta(\sin^2\theta_{23})$, since it is very strongly dependent on the input value, we avoid discussing it.

\subsection{Establishing CPT-Violation}
\label{subsec:MeasViolation}

If neutrino and antineutrino oscillation parameters are indeed different, it is possible that DUNE and Hyper-K will be able to establish that CPT is not an exact symmetry. Here, we simulate data assuming that neutrino and antineutrino parameters are different. We choose input values that are consistent with the current world's neutrino data. We will restrict the discussion to the hypothesis that the parameters $(\theta_{13}, \delta)$ and $(\overline{\theta}_{13}, \overline{\delta})$ are not equal. The motivation is as follows. While the Daya Bay experiment places a strong constraint on the value of $\sin^2{\left(2\overline{\theta}_{13}\right)} = 0.0841 \pm 0.0033$, as discussed in Section~\ref{subsubsec:NuBarParams}, the results from NO$\nu$A  and T2K  point to much higher values for the parameter $\sin^2{\left(2\theta_{13}\right)}$ -- as high as $0.2$ for particular values of $\delta_{CP}$. On the other hand, $\delta_{CP}$ and $\overline{\delta}_{CP}$ are only, at best, poorly constrained. 

We consider two CPT-violating scenarios -- Scenario A and Scenario B. In both scenarios, $\Delta m^2_{31}=\Delta \overline{m}^2_{31}=2.524\times 10^{-3}$~eV$^2$ and $\sin^2\theta_{23}=\sin^2\overline{\theta}_{23}=0.500$, and the inputs for $\Delta m^2_{21}$, $\Delta \overline{m}^2_{21}$, $\sin^2\theta_{12}$, $\sin^2\overline{\theta}_{12}$ are set to their best fit values, discussed in  Sections~\ref{subsubsec:NuBarParams} and \ref{subsubsec:NuParams}, and listed in Tables~\ref{tab:InputParamsScenarioA} and \ref{tab:InputParamsScenarioB}. The fact that these long-wavelength parameters are also CPT-violating has no significant effect on the results discussed below. 

In Scenario A, the mixing angle $\theta_{13}$ is identical for neutrinos and antineutrinos, but the CP-violating phases $\delta_{CP}$ and $\overline{\delta}_{CP}$ are maximally different, as listed in Table~\ref{tab:InputParamsScenarioA}.
\begin{table}[!htbp]
\begin{center}
\begin{tabular}{|r||c|c|c|c|}
\hline
Neutrino Parameter & $\sin^2{\left(2\theta_{13}\right)}$ & $\delta_{CP}$ & $\sin^2\theta_{12}$ & $\Delta m_{21}^2$ \\ \hline
Value & $0.084$ & $-\pi/2$ & $0.311$ & $5.4\times 10^{-5}$ eV$^2$ \\ \hline \hline
Antineutrino Parameter & $\sin^2{\left(2\overline{\theta}_{13}\right)}$ & $\overline{\delta}_{CP}$ & $\sin^2\overline{\theta}_{12}$ & $\Delta \overline{m}_{12}^2$ \\ \hline
Value & $0.084$ & $\pi/2$ & $0.342$ & $7.58\times 10^{-5}$ eV$^2$ \\ \hline
\end{tabular}
\end{center}
\caption{Parameters taken as physical inputs for the CPT-violating Scenario A. Unlisted parameters are the same as in Table~\ref{tab:InputParamsExclusion}, with $\sin^2\theta_{23}=\sin^2\overline{\theta}_{23}=0.500$.}
\label{tab:InputParamsScenarioA}
\end{table}
\begin{table}[!htbp]
\begin{center}
\begin{tabular}{|r||c|c|c|c|}
\hline
Neutrino Parameter & $\sin^2{\left(2\theta_{13}\right)}$ & $\delta_{CP}$ & $\sin^2\theta_{12}$ & $\Delta m_{21}^2$ \\ \hline
Value & $0.2$ & $\pi/2$ & $0.311$ & $5.4\times 10^{-5}$ eV$^2$ \\ \hline \hline
Antineutrino Parameter & $\sin^2{\left(2\overline{\theta}_{13}\right)}$ & $\overline{\delta}_{CP}$ & $\sin^2\overline{\theta}_{12}$ & $\Delta \overline{m}_{12}^2$ \\ \hline
Value & $0.084$ & $\pi/2$ & $0.342$ & $7.58\times 10^{-5}$ eV$^2$ \\ \hline
\end{tabular}
\end{center}
\caption{Parameters taken as physical inputs for the CPT-violating Scenario B. Unlisted parameters are the same as in Table~\ref{tab:InputParamsExclusion}, with $\sin^2\theta_{23}=\sin^2\overline{\theta}_{23}=0.500$.}
\label{tab:InputParamsScenarioB}
\end{table}
In Scenario B, the mixing angles $\theta_{13}$ and $\overline{\theta}_{13}$ are different, but the CP-violating phases are the same. The largest difference between the Daya Bay measurement of $\sin^2\overline{\theta}_{13}$ and the NO$\nu$A and T2K measurements of $\sin^2\theta_{13}$ occurs for $\delta_{CP} = \pi/2$, so we choose this value for $\delta_{CP}$  and $\overline{\delta}_{CP}$, as listed in Table~\ref{tab:InputParamsScenarioB}.

\begin{figure}
\begin{center}
\includegraphics[width=\linewidth]{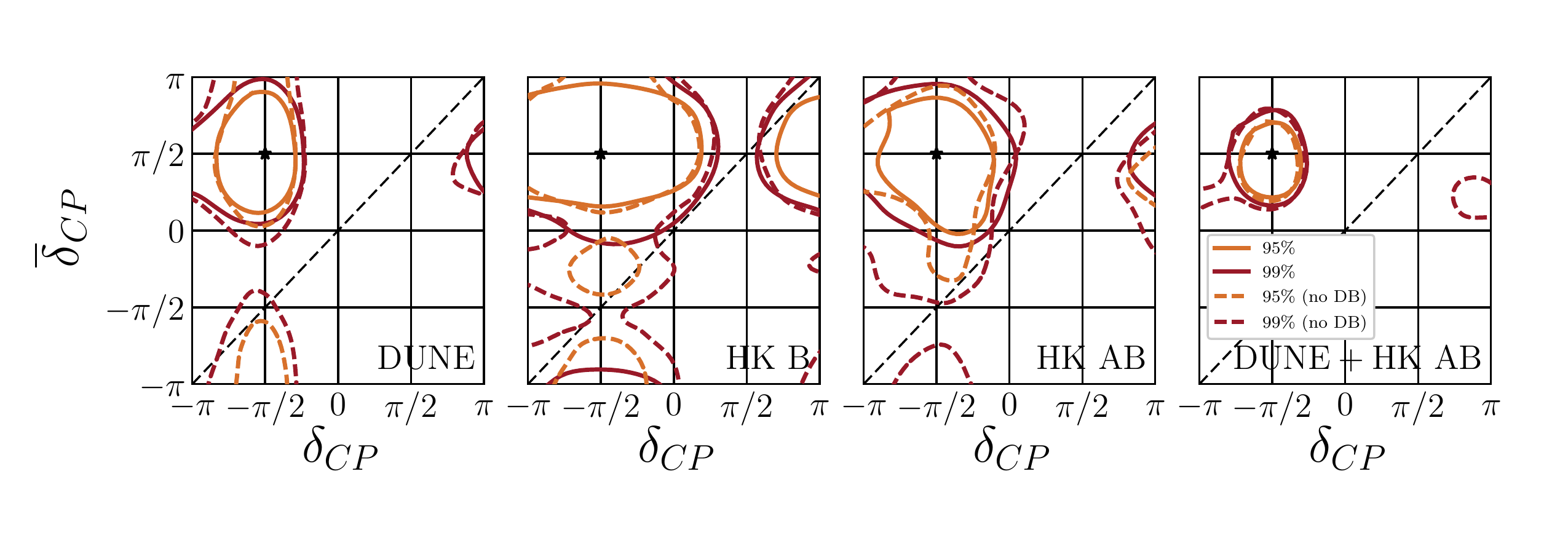}
\caption{Results of the measurement of the parameters $\delta_{CP}$ and $\overline{\delta}_{CP}$ in the CPT-violating Scenario A, in which the values of these two parameters are maximally different. Each panel represents the measurement from a particular combination of experiments, with all ten unseen parameters marginalized. Shown are the 95\% (orange), and 99\% (red) credible regions for the two parameters. Solid lines display the results for the analysis including the prior on $\sin^2(2\overline{\theta}_{13})$ from the Daya Bay experiment, and dotted lines display results without this prior.}
\label{fig:ResultsScenarioA}
\end{center}
\end{figure}

\begin{figure}
\begin{center}
\includegraphics[width=\linewidth]{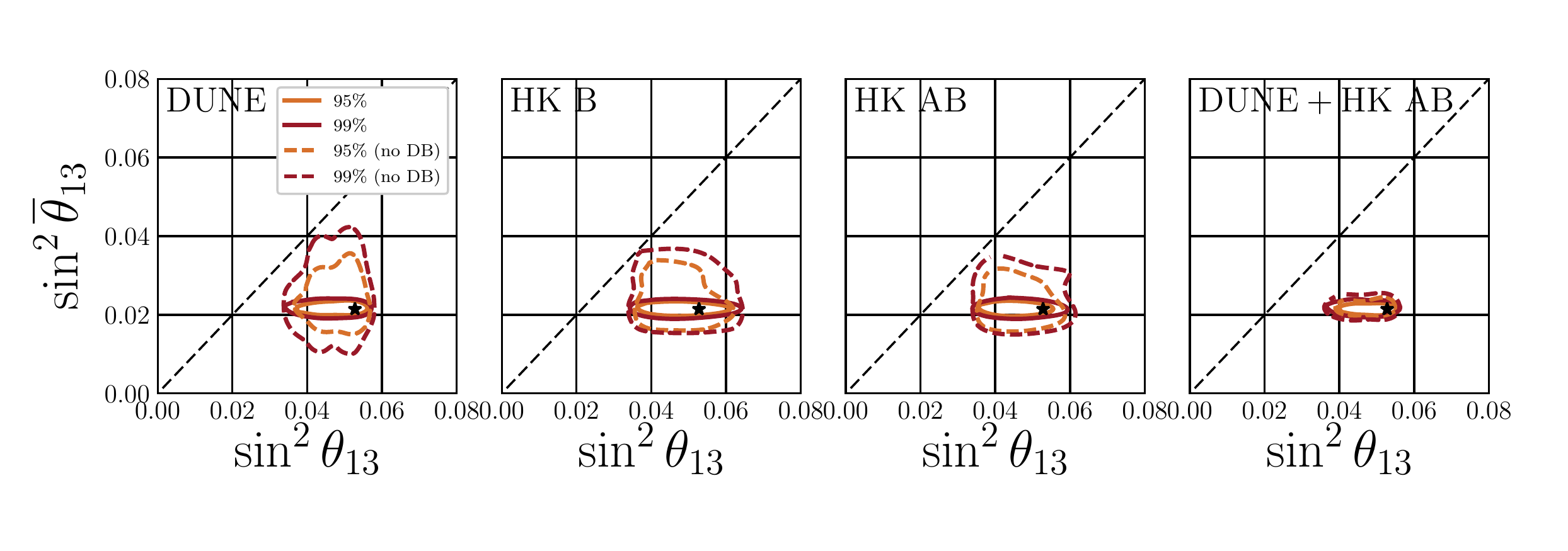}
\caption{Results of the measurement of the parameters $\sin^2\theta_{13}$ and $\sin^2\overline{\theta}_{13}$ in the CPT-violating Scenario B. Each panel represents the measurement from a particular combination of experiments, with all ten unseen parameters marginalized. Shown are the 95\% (orange), and 99\% (red) credible regions for the two parameters. Solid lines display the results for the analysis including the prior on $\sin^2(2\overline{\theta}_{13})$ from the Daya Bay experiment, and dotted lines display results without this prior.}
\label{fig:ResultsScenarioB}
\end{center}
\end{figure}

For each Scenario, we proceed as in Section~\ref{subsec:Exclusion}, including all of the same combinations of the DUNE and Hyper-K simulated data. Each analysis is repeated twice, once with the prior on $\sin^2(2\overline{\theta}_{13})$ from the Daya Bay experiment and once without. Fig.~\ref{fig:ResultsScenarioA} depicts the extracted allowed regions of the $\delta_{CP}\times \overline{\delta}_{CP}$ plane, marginalized over all other oscillation parameters, for Scenario A. Different-color contours correspond to $95\%$ and $99\%$ credibility while the solid (dashed) contours correspond to analyses performed with (without) the Daya Bay prior, Eq.~(\ref{eq:DB}). Fig.~\ref{fig:ResultsScenarioB} depicts the extracted allowed regions of the $\sin^2\theta_{13}\times \sin^2\overline{\theta}_{13}$-plane, marginalized over all other oscillation parameters, for Scenario B. Different-color contours correspond to $95\%$ and $99\%$ credibility while the solid (dashed) contours correspond to analyzes perform with (without) the Daya Bay prior, Eq.~(\ref{eq:DB}).
%
%
%
%
In both figures, the diagonal dashed line corresponds to the CPT-conserving region of the parameter subspace.

Fig.~\ref{fig:ResultsScenarioA} reveals that, if nature agrees with Scenario A, the DUNE experiment, by itself, could ``discover'' CPT-violation by establishing that $\delta_{CP}\neq\overline{\delta}_{CP}$ at more than the 99\%~credibility level, especially if existing results from Daya Bay are included in the analysis. The case for CPT-violation would be even stronger if one were to combine DUNE data with Hyper-K beam and atmospheric data. This discrepancy would exceed the existing 2$\sigma$ discrepancy between $\Delta m^2_{21}$ and $\Delta \overline{m}^2_{21}$, assuming it does not change significantly with future data from JUNO or RENO-50.  

Fig.~\ref{fig:ResultsScenarioB} reveals that, if nature agrees with Scenario B, the Hyper-K or DUNE experiments, by themselves, could ``discover'' CPT-violation by establishing that $\theta_{13}\neq\overline{\theta}_{13}$ at more than the 99\%~credibility level. Here the impact of including existing results from Daya Bay in the analysis is most transparent. The case for CPT-violation would be  stronger if one were to combine DUNE data with Hyper-K beam and atmospheric data. Once all experimental data are combined, the measurement power gained by including prior information from Daya Bay has largely vanished. This discrepancy would also significantly exceed the existing $2\sigma$ discrepancy between $\Delta m^2_{21}$ and $\Delta \overline{m}^2_{21}$, assuming it does not change significantly with future data from JUNO or RENO-50.  


\setcounter{footnote}{0}
\setcounter{equation}{0}
\section{Discussion and Conclusions}
\label{sec:Conclusions}

We explore how well the next-generation long-baseline experiments DUNE and Hyper-K can test the hypothesis that neutrino and antineutrino oscillation parameters -- mixing angles, mass-squared differences, CP-odd phases -- are identical, as dictated by the CPT-theorem. We first discussed the current status of neutrino versus antineutrino oscillation parameters, highlighting the fact that order 100\% differences are not excluded by existing neutrino data for all oscillation parameters. 

Currently, the long-wavelength parameters -- $\Delta m^2_{21}$, $\sin^2\theta_{12}$ and their antineutrino counterparts -- are constrained by comparing solar neutrino data with those from the KamLAND experiment. $\Delta(\Delta m^2_{21})$ deviates from zero at the $2\sigma$ level, while $\Delta(\sin^2\theta_{12})$ values as large as 0.4 are safely allowed \cite{deGouvea:2004va}. Neither Hyper-K nor DUNE (nor combinations of the two) are expected to modify the current situation. On the other hand, next-generation long-baseline reactor experiments, like JUNO \cite{An:2015jdp} and RENO-50  \cite{Kim:2014rfa}, will measure the antineutrino parameters $\sin^2\overline{\theta}_{12}$ and $\Delta \overline{m}^2_{21}$ much more precisely. 
These are expected to have a limited impact on $\Delta(\Delta m^2_{21})$ given that the dominant contribution to its uncertainty comes from current solar data, unless the preferred value differs significantly from the one reported by KamLAND. $\Delta(\sin^2\theta_{12})$ can change significantly if JUNO/RENO-50 have the ability to tell whether $\overline{\theta}_{12}$ is in the 
``light'' or ``dark'' side (or $\cos2\overline{\theta}_{12}$ is positive or negative). Estimating in detail whether this is the case is beyond the ambitions of this manuscript and will be left for possible future exploration. Here we can, however, state that determining the sign of $\cos2\overline{\theta}_{12}$ and the antineutrino mass-hierarchy via $\bar{\nu}_e$-disappearance in vacuum is not possible. However, if the antineutrino mass-hierarchy -- whether $\Delta \overline{m}^2_{32}$ is positive or negative -- were established independently by a different experiment, it stands to reason that JUNO/RENO-50 should be as sensitive to the sign of $\cos2\overline{\theta}_{12}$ as it is sensitive to determining the mass-hierarchy in the CPT-conserving hypothesis, when one can take advantage of the fact that $\cos2\theta_{12}$ is known to be positive from solar neutrino data. 

$\Delta(\Delta m^2_{32})$ is also currently allowed to be as large as -- in fact double --  the parameters themselves, around $5\times 10^{-3}$ eV$^2$. The reason is that the neutrino and antineutrino mass-hierarchies are not know so it is possible that, for example, $\Delta \overline{m}^2_{32}<0$ while  $\Delta m^2_{32}>0$. DUNE and Hyper-K will change the situation qualitatively. If CPT is a good symmetry, we expect these experiment will be able to constrain $\Delta(\Delta m^2_{32})$ to values, in magnitude, smaller than $5\times 10^{-5}$ eV$^2$. This is possible only because the neutrino and antineutrino mass-hierarchies can be independently determined. DUNE can do it without the help of Hyper-K, as long as current measurements on $\sin^2\overline{\theta}_{13}$ from Daya Bay, RENO, and Double Chooz, are taken into account. On the other hand, DUNE combined with Hyper-K, including the very large atmospheric muon-type neutrino sample, can significantly constrain $\Delta(\Delta m^2_{32})$ even if one ignores information from the existing reactor antineutrino experiments. 

$\Delta(\sin^2\theta_{13})$ and $\Delta(\sin^2\theta_{23})$ are also currently allowed to be as large as the parameters themselves. Similar to $\Delta(\sin^2\theta_{12})$, $\Delta(\sin^2\theta_{23})$ suffers from the octant-degeneracy. We did not fully explore this issue here but have determined that resolving the octant degeneracy in the neutrino sector may prove to be very challenging and, perhaps, outside the reach of DUNE and Hyper-K. The precision on $\Delta(\sin^2\theta_{13})$ is currently limited by the capabilities of the current long-baseline experiments to measure the neutrino oscillation parameter $\sin^2\theta_{13}$. The situation is expected to improve significantly with data from DUNE and Hyper-K. Finally, while $\Delta(\delta_{CP})$ is currently unconstrained, DUNE and Hyper-K have the ability to measure CP-violation in the neutrino and antineutrino sectors independently, especially if existing information on $\sin^2\overline{\theta}_{13}$ is included in the fit. 

We also explored whether DUNE and Hyper-K can discover CPT violation if it manifests itself as $\delta_{CP}\neq\overline{\delta}_{CP}$ or $\sin^2\theta_{13}\neq\sin^2\overline{\theta}_{13}$. For the choice of parameters we considered, the answer is positive, as depicted in Figs.~\ref{fig:ResultsScenarioA} and \ref{fig:ResultsScenarioB}.

It is sensible to ask whether any of these current or future bounds represent very strong tests of the CPT-theorem. Numerically, none on the bounds on mixing parameters are expected to be very stringent. Furthermore, while the bounds on the differences between mass-squared differences are potentially rather stringent, they do not constrain the possibility that the differences between neutrino and antineutrino masses -- as opposed to mass-squared differences -- are relatively large. For example the hypothesis $m_1=0.000001$~eV, $\overline{m}_1=0.01$~eV, $\Delta m^2_{ij}=\Delta \overline{m}^2_{ij},(\forall i,j)$, cannot be tested by neutrino oscillation experiments. Nonetheless, the comparisons discussed here are unique tests of the CPT-theorem and hence provide irreplaceable information on fundamental physics.  

The results presented and discussed here also illustrate some of the challenges of testing the three-massive-neutrinos paradigm. Regardless of whether the CPT-theorem is in question, in order to test the three-massive-neutrinos paradigm in a way that is as model-independent as possible, it is crucial to measure the neutrino oscillation parameters using distinct probes: low-energy neutrinos versus high-energy neutrinos, neutrinos from the Sun versus neutrinos from nuclear reactors versus neutrinos from pion decay, neutrinos versus antineutrinos, etc. We have explored here the capabilities and limitations of ``superbeam'' experiments to independently consider neutrino and antineutrino oscillations. We find that, for example, many antineutrino measurements can only be done with precision if external antineutrino data from nuclear reactors is also included. Given that the number of accessible neutrino observables is limited, access to different neutrino sources -- accelerators, the Sun, nuclear reactors, etc -- detector-types, and propagation media -- vacuum versus dense matter -- is necessary if we are to move significantly towards exploring the hypothesis that there is more new physics in the leptonic sector than nonzero neutrino masses. 

\section*{Acknowledgements}
This work is supported in part by DOE grant \#de-sc0010143. We acknowledge the use of the Quest computing cluster at Northwestern University for a portion of this research.

\bibliographystyle{apsrev-title}
\bibliography{CPTBib}{}

\end{document}